\documentclass[aps,prd,nofootinbib,preprintnumbers,onecolumn]{revtex4-2} 
\usepackage{amsmath,amssymb}
\usepackage{graphicx}  
\usepackage{bbold}
\usepackage{slashed}
\newcommand{\bL}{\begin{Large}}
\newcommand{\eL}{\end{Large}}

\newcommand{\bea}{\begin{eqnarray}}
\newcommand{\eea}{\end{eqnarray}}
\newcommand{\be}{\begin{equation}}
\newcommand{\ee}{\end{equation}}

\usepackage[colorinlistoftodos]{todonotes}
\usepackage{appendix}

\DeclareGraphicsExtensions{.pdf,.png,.jpg,.eps}
\graphicspath{ {./}{./figs/} }

\begin{document}

\preprint{\rightline{KCL-PH-TH/2025-{\bf 35}}}

\title{Quantum Gravity and Entanglement in Particle Physics and Gravitation}

\author{Nick E. Mavromatos$^{a,b}$}

\medskip
\affiliation{$^a$Physics Division, School of Applied Mathematical and Physical Sciences, National Technical University of Athens, Zografou Campus, Athens 157 80, Greece}
\affiliation{$^b$Theoretical Particle Physics and Cosmology Group, Department of Physics, King's College London, London, WC2R 2LS, UK}

\date{\today}

\begin{abstract}
Some approaches to Quantum Gravity (QG) entail decoherence of quantum matter propagating in it, due to an ``environment'' of QG degrees of freedom inaccessible to low-energy observers.
 In the first part of this talk, I discuss potential, and rather unique, effects of QG-induced decoherence on entangled particle states, specifically an induced modification of Einstein-Podolsky-Rosen (EPR) correlations of entangled neutral-meson states in meson factories ($\omega$-effect). In the second part, I summarise a recent work in which axion-like fields, forming a kind of condensate clouds surrounding rotating (Kerr-type) astrophysical black holes, can lead to superradiant instabilities, and, through these, to the production of EPR-like entangled states of gravitons, with the entanglement pertaining to  (left, right) polarisation degrees of freedom. In the presence of axions and Kerr geometries, there are non-trivial gravitational Chern-Simons (gCS)-type anomalous terms in the respective low-energy gravitational effective actions. Depending on whether the graviton entanglement is due to the non-anomalous terms (of General-Relativity type) in the effective action, or to the gCS terms, one obtains different structures of the resulting entangled squeezed-graviton states, which resemble somewhat the $\omega$-effect.

\end{abstract}


\maketitle

\section{Introduction: approaches to Quantum Gravity}	

This year celebrates the 100th anniversary of the birth of Quantum Theory~\cite{Heisenberg:1925zz}, whose ``characteristic trait'' is {\it entanglement}, according to Schr\"odinger's interpretation~\cite{Schrodinger:1935zz} of the Einstein-Podolsky-Rosen (EPR) ``paradox''~\cite{epr}. Entanglement has proven an important property of quantum mechanics, which has led to a plethora of breakthrough applications in quantum information and quantum computation, including multipartite entanglement~\cite{Kwiat:1995uur,Pan:1998zdc,Bennett:2000zoh,Nielsen:2012yss,Toth:2012lpv,Pan:2012svn,Hiesmayr:2017xgx}. It has 
always been an important ingredient of particle physics~\cite{Shi:2025ggs}, which is based on quantum field theory.
Intense studies of entangled neutral mesons in meson factories have been performed for some time,\cite{Amelino-Camelia:2010cem,KLOE:2010yad,BaBar:2014omp,KLOE-2:2021ila} providing stringent tests of quantum mechanics \cite{Hiesmayr:2011na}. Recently, entanglement has been observed in LHC experiments, in $t {\overline t}$ pairs~\cite{ATLAS:2023fsd,CMS:2024pts,CMS:2024zkc} (see related talks in this conference).

Also, this year marks the 110th  anniversary of the discovery of General Relativity (GR)~\cite{Einstein:1915by}, the classical field theory of Gravitation. Yet, Quantum Gravity (QG), that is a quantisation of the gravitational interaction, remains an elusive concept. QG is  supposed to be a quantum theory that explains the dynamical emergence of spacetime itself from more fundamental structures. At present, there is no experimental evidence, not even hints, for QG, although there are several, but not complete, theoretical approaches, some background dependent, 
such as string theory~\cite{str} and asymptotic safety~\cite{as,as2}), which can incorporate the Standard Model in their low-energy formalism, 
others background independent, 
such as loop QG~\cite{loop,loop2} and quantum foam~\cite{Oriti:2001qu,Asante:2020qpa},which, however, are phenomenologically less advanced.\footnote{Some physicists believe that Gravity may not even be a fundamental interaction, but rather an entropic force, associated with information carried by the position of material bodies, in an appropriate holographic setting~\cite{Verlinde}.} 
In more phenomenological settings, including also versions of (non-critical) string theory and its brane~\cite{Polchinski:1996na} extensions~\cite{aemn1,Ellis:2008gg}, 
there are also models of QG that violate~\cite{Ellis:1999rz,Ellis:1999uh} or deform~\cite{Amelino-Camelia:2002cqb,Magueijo:2001cr} Lorentz symmetry, subject to a plethora of tests using cosmic multimessenger probes~\cite{aemns,efmmn,Addazi:2021xuf}.  Such models may entail a violation of CPT symmetry, given that one of the basic assumptions of the CPT theorem~\cite{Schwinger:1951xk,Pauli_CPT,Bell:1955djs,Luders:1957bpq}, Lorentz invariance, is violated.\footnote{In Ref.~\cite{Greenberg:2002uu} a sort of an inverse of CPT theorem has also been presented, using concepts from scattering theory, according to which violation of CPT implies violation of Lorentz invariance, provided however that the transfer matrix is well defined.} 
In such cases, the CPT generator is a well-defined quantum mechanical operator, not commuting with the Hamiltonian, $H$, of the low-energy system.

However, QG may entail another, highly non-trivial, property, that of inducing {\it decoherence} of (low-energy) quantum matter (and radiation) propagating in it, given that there might be an ``environment'' of QG degrees of freedom  (microscopic (Planck size) black holes, and other topologically non-trivial configurations) inaccessible to low-energy observers, who perform scattering experiments for their observations. In the presence of QG decoherence, the scattering matrix is not well defined~\cite{Hawking:1982dj} in the effective low-energy local field theory, whose time evolution is assumed, in a phenomenological setting~\cite{Ellis:1983jz,Banks:1983by}, to obey Gorini-Kossakowski--Lindblad-Sudarshan-type equations~\cite{Gorini:1975nb,Lindblad:1975ef} for the respective density matrices.

Wald has argued~\cite{Wald:1980nm} that in such QG-induced decoherening situations, the CPT operator in the effective theory of low-energy matter probes is {\it not well-defined}.\footnote{In this case, the application of the theorem of Ref.~\cite{Greenberg:2002uu} is unclear, since a transfer matrix is not well defined. Hence, definite conclusions on the Lorentz (non)invariance of the splitting between subsystem and QG environment cannot be reached \cite{Milburn:2003zj}.} This is an {\it intrinsic CPT Violation}, which has ``smoking-gun'' consequences for the EPR correlations of entangled particle states ($\omega$-effect~\cite{Bernabeu:2003ym}). This effect is distinct from the case where CPT is violated due to Lorentz-symmetry violation in, say, SM extensions~\cite{Kostelecky:2008ts}, in which the CPT operator is well-defined, but non-commuting with the Hamiltonian of the system. 

The structure of the talk is the following: in the next section \ref{sec:omega}, we discuss the $\omega$-effect, and its current experimental bounds and theoretical estimates. In section \ref{sec:superrad}, we discuss the emergence of entangled (non-separable) graviton-polarisation states from axion-like fields in superradiant clouds of rotating astrophysical black holes~\cite{Dorlissuperr,Dorlis:2025amf}. We also draw some remote analogies with the $\omega$-effect. Finally, section \ref{sec:concl} gives our conclusions and outlook. 

\vspace{-0.2cm}

\section{QG modifications of EPR Correlations of entangled neutral mesons - The $\omega$-effect}\label{sec:omega}

In Ref.~\cite{Bernabeu:2003ym}, it was suggested that the aforementioned ill-defined nature of the CPT operator would imply modifications of the EPR correlations by contaminating - with the ``wrong-symmetry state''- the entangled state of the two neutral mesons produced in a meson factory via the decay of another appropriate neutral meson (e.g. a $\Phi$ particle in the case of Kaons in $\Phi$-factories~\cite{Amelino-Camelia:2010cem}): 
\begin{align}\label{omega}
\left|  i\right\rangle  &  =\frac{1}{\sqrt{2}}\left(  \left|  \overline{M_{0}%
}\left(  \overrightarrow{k}\right)  \right\rangle \left|  M_{0}\left(
-\overrightarrow{k}\right)  \right\rangle -\left|  M_{0}\left(
\overrightarrow{k}\right)  \right\rangle \left|  \overline{M_{0}}\left(
-\overrightarrow{k}\right)  \right\rangle \right) \nonumber \\
&  +\frac{\omega}{\sqrt{2}}\left(  \left|  \overline{M_{0}}\left(
\overrightarrow{k}\right)  \right\rangle \left|  M_{0}\left(  -\overrightarrow
{k}\right)  \right\rangle +\left|  M_{0}\left(  \overrightarrow{k}\right)
\right\rangle \left|  \overline{M_{0}}\left(  -\overrightarrow{k}\right)\,.
\right\rangle \right)
\end{align}
In the above formula, $\vert M_0 \rangle$ denotes the ``flavour'' state of the neutral mesons, and the overline corresponds to the antiparticle state, while 
$\omega=\left|  \omega\right|  e^{i\Omega}$ is a complex CPT violating
(CPTV) parameter, parametrising the effects of the ill-definition of the generator of the CPT symmetry due to QG decoherence. In the absence of such a decoherence ({\it i.e.} when $\omega=0$), in which case the CPT operator can
be defined as a quantum mechanical operator, even if it does not commute with the Hamiltonian of the entangled neutral mesons, the decay of a (generic) initial 
meson with quantum numbers $J^{{\mathcal P} C} = 1^{ - - } $ (where ${\mathcal P}$ permutes spatial coordinates and $C$ is the charge conjugation)~\cite{Lipkin:1968ygs}, leads to
a pair state of neutral mesons $\left|  i\right\rangle $ having the form of an entangled state that has the Bose symmetry associated with 
particle-antiparticle indistinguishability.
The entangled state would vanish for identical bosons. However, in case CPT is ill-defined due
to space-time foam, then $M_{0}$ and $\overline{M_{0}}$ are not identical, and this leads to the modified EPR  entangled state \eqref{omega} (this is the ``$\omega$-effect''.\cite{Bernabeu:2003ym}). On concentrating, from now on for concreteness, on the case of neutral Kaons in $\Phi$ factories, we remark that,  in terms of mass eigenstates, $\vert K_{L,S} \rangle$, the initial state \eqref{omega} reads \cite{Bernabeu:2003ym}: 
\begin{align}\label{CPTV}
\left|  i\right\rangle &=\mathcal{C}\,
\Big[
\left(  \left|  K_{L}\left(  \overrightarrow{k}\right)  \right\rangle \left|
K_{S}\left(  -\overrightarrow{k}\right)  \right\rangle -\left|  K_{S}\left(
\overrightarrow{k}\right)  \right\rangle \left|  K_{L}\left(  -\overrightarrow
{k}\right)  \right\rangle \right)  \nonumber \\
&+\, \omega\left(  \left|  K_{S}\left(  \overrightarrow{k}\right)  \right\rangle
\left|  K_{S}\left(  -\overrightarrow{k}\right)  \right\rangle -\left|
K_{L}\left(  \overrightarrow{k}\right)  \right\rangle \left|  K_{L}\left(
-\overrightarrow{k}\right)  \right\rangle \right)
\Big]\,,
\end{align}
where $\mathcal C$ is an appropriate normalization factor, which depends on the CP-violating parameters; its explicit form is of no relevance to our considerations here. 

In Ref.~\cite{Bernabeu:2006av}, a two-level ``spin''-system analogue of the $\omega$-effect has been described.
In the notation of two-level systems (on suppressing the $\overrightarrow{k}$
label) we can write
$\left|  K_{L}\right\rangle    =\left|  \uparrow\right\rangle \, , 
\left|  K_{S}\right\rangle    =\left|  \downarrow\right\rangle$.
In this model, the quantum spacetime foam is represented by (quantum fluctuating) compactified brane spacetime defects (``D-foam'') \cite{Ellis:2005ib}, which interact with the low-energy matter~\cite{Ellis:2008gg}, represented by open strings with their ends attached to a D3-brane world, corresponding to our Universe. Such interactions back-react onto the (3+1)-dimensional spacetime, leading to an induced ``recoil'' metric, allowing ``spin-flip'', and undergoing stochastic random fluctuations, representing the ``foamy'' nature of quantum spacetime. The upshot is an effective interaction Hamiltonian for the two-state system, whose dominant features for an induced $\omega$-effect are captured by the term \cite{Bernabeu:2006av}:
\begin{equation}
\widehat{H_{I}} = -\left(  { r_{1} \sigma_{1} + r_{2} \sigma_{2}} \right)
\widehat{k} \label{inthamil}
\end{equation}
to leading order in the small stochastic parameters $r_i\,, i=1.2,$, with $
\left\langle r_{i}\right\rangle =0,\;\left\langle r_{i}r_{j}\right\rangle =\Delta_{i}\delta_{ij}$, $i,j=1.2$.
The ``gravitationally-dressed'' meson energy eigenstates in such a case are given, in standard perturbation theory, by:
{\small \begin{equation}
\left|  k^{\left(  i\right)  },\downarrow\right\rangle _{QG}^{\left(
i\right)  } = \left|  k^{\left(  i\right)  },\downarrow\right\rangle ^{\left(
i\right)  } + \left|  k^{\left(  i\right)  },\uparrow\right\rangle ^{\left(
i\right)  } \alpha^{\left(  i\right)  }, \qquad \alpha^{\left(  i\right)  }= \frac{^{\left(  i\right)  }\left\langle \uparrow,
k^{\left(  i\right)  }\right|  \widehat{H_{I}}\left|  k^{\left(  i\right)  },
\downarrow\right\rangle ^{\left(  i\right)  }}{E_{2} - E_{1}}\,.
\label{qgpert}
\end{equation}}
The dressed state corresponding to $\left|  { k^{\left(  i\right)  }, \uparrow}
\right\rangle ^{\left(  i \right)  } $ is obtained from
\eqref{qgpert} by $\left|  \downarrow\right\rangle \leftrightarrow\left|
\uparrow\right\rangle $ and $\alpha\to\beta$ where
$\beta^{\left(  i\right)  }= \frac{^{\left(  i\right)  }\left\langle
\downarrow, k^{\left(  i\right)  }\right|  \widehat{H_{I}}\left|  k^{\left(
i\right)  }, \uparrow\right\rangle ^{\left(  i\right)  }}{E_{1} - E_{2}}$.
Here the quantities $E_{i} = (m_{i}^{2} + k^{2})^{1/2}$ denote the energy
eigenvalues, and $i=1$ is associated with the up state and $i=2$ with the down
state. In Ref.~\cite{Bernabeu:2006av}, it was postulated that the initial 
totally antisymmetric boson state in a meson factory is a perturbed, ``gravitationally-dressed'' antisymmetric state, which can be expressed in terms of the unperturbed single-particle states as:
\begin{align}
&\left|  {k, \uparrow} \right\rangle _{QG}^{\left(  1 \right)  } \left|  { - k,
\downarrow} \right\rangle _{QG}^{\left(  2 \right)  } - \left|  {k,
\downarrow} \right\rangle _{QG}^{\left(  1 \right)  } \left|  { - k, \uparrow}
\right\rangle _{QG}^{\left(  2 \right)  } = \nonumber \\
&\left|  {k, \uparrow} \right\rangle ^{\left(  1 \right)  } \left|  { - k,
\downarrow} \right\rangle ^{\left(  2 \right)  } - \left|  {k, \downarrow}
\right\rangle ^{\left(  1 \right)  } \left|  { - k, \uparrow} \right\rangle
^{\left(  2 \right)  } \nonumber \\
&+ \left|  {k, \downarrow} \right\rangle ^{\left(  1 \right)  } \left|  { - k,
\downarrow} \right\rangle ^{\left(  2 \right)  } \left(  {\beta^{\left(  1
\right)  } - \beta^{\left(  2 \right)  } } \right)  + \left|  {k, \uparrow}
\right\rangle ^{\left(  1 \right)  } \left|  { - k, \uparrow} \right\rangle
^{\left(  2 \right)  } \left(  {\alpha^{\left(  2 \right)  } - \alpha^{\left(
1 \right)  } } \right) \nonumber \\
&+ \beta^{\left(  1 \right)  } \alpha^{\left(  2 \right)  } \left|  {k,
\downarrow} \right\rangle ^{\left(  1 \right)  } \left|  { - k, \uparrow}
\right\rangle ^{\left(  2 \right)  } - \alpha^{\left(  1 \right)  }
\beta^{\left(  2 \right)  } \left|  {k, \uparrow} \right\rangle ^{\left(  1
\right)  } \left|  { - k, \downarrow} \right\rangle ^{\left(  2 \right)  }\,.
\label{entangl}
\end{align}
Conservation of strangeness characterises the $\omega$-effect of \cite{Bernabeu:2003ym}, relevant for $\Phi$-factories~\cite{Amelino-Camelia:2010cem}, which in the above model amounts to setting $r_i \propto \delta_{i2}$, for which $\alpha^{(i)}=\beta^{(i)}$. On the other hand, for $r_i \propto \delta_{i1}$, one has $\alpha^{(i)}=-\beta^{(i)}$, and the initial state of neutral Kaons 
\eqref{omega} has no definite strangeness. QG does not have to conserve this quantum number, which is reflected in the above spin-model.\footnote{In Ref.~\cite{Bernabeu:2006av} the generation of an $\omega$-effect via time evolution, in the presence of the interaction \eqref{inthamil} has been discussed, with the result that a strangeness-conserving initial state leads, under temporal evolution in the foam, to the appearance of CPT violating terms with a 
strangeness-violating form, while an initially strangeness-violating 
combination generates a 
strangeness-conserving $\omega$-effect~\cite{Bernabeu:2003ym}.} 
The propagation of the meson eigenstates in a D-particle foam leads to an estimate of the $\omega$ effect~\cite{Bernabeu:2006av}:
\begin{equation}
|\omega|^{2} = \mathcal{O}\left(  \frac{(\langle
\downarrow, k |H_{I} |k, \uparrow\rangle)^{2}
}
{(E_{1} - E_{2})}  \right)  \sim \frac{\Delta_{2}
k^{2}}{(m_{1} - m_{2})^{2}} \label{omegaorder}%
\end{equation}
for the physically interesting case of meson-factories in which the momenta are of order of the rest masses. The variance 
$\Delta_{2} = \zeta^{2} k^{2}/M_{\rm QG}^{2}$, where
$\zeta$ is at present a phenomenological parameter, indicating the momentum transfer during the interactions of low-energy matter with the spacetime foam. This can only be determined in an UV complete theory of QG. In the D-foam case~\cite{Ellis:2008gg}, the QG scale 
$M_{\rm QG} = M_s / n^\star$
where $n^\star$ is the linear density of D-foam compactified brane defects encountered  in the propagation of the mesons, and $M_s$ is the string scale, which is in general different from the (reduced) Planck scale $M_{\rm Pl} = 2.435 \times 10^{18}~\rm GeV$. 

The most stringent bounds to date of the $\omega$ effect have been provided by KLOE-2 Collaboration~\cite{KLOE-2:2021ila,DiDomenico:2023kbu} in the upgraded DA$\Phi$NE $\Phi$-factory \cite{Amelino-Camelia:2010cem}:
\begin{align}\label{kloe2omega}
{\rm Re}\omega &= \Big(-2.3^{+1.9}_{-1.5} \pm 0.7 \Big) \times 10^{-4}\,, \quad {\rm Im}\omega = \Big(-4.1^{+2.8}_{-2.6} \pm 4.9 \Big) \times 10^{-4}\,, \nonumber \\ |\omega| &< \Big(4.7 \pm 2.9 \pm 1.0 \Big) \times 10^{-4}\,,
\end{align}improving earlier results by KLOE Collaboration~\cite{KLOE:2006iuj,KLOE:2010yad}. The $\Phi$ factories are the most sensitive probe of the $\omega$-effect, due to the fact that, in the relevant asymmetry 
$I(\Delta t) = \frac{1}{2} \int_{\vert \Delta t \vert }^{\infty}\, \vert A(\pi^+\pi^-, \pi^+\pi^- )\vert ^2 $ of the $\pi^+\pi^-$ decay channels of the neutral Kaons, the parameter $\omega$ appears in the combination~\cite{Bernabeu:2003ym} $\vert \omega \vert/\vert \eta_{+-}\vert$, where the CP-violation parameter $\vert \eta_{+-}\vert = \mathcal O(10^{-3})$, as measured in the experiment.
Recently, weaker but independent bounds for $\vert \omega \vert $, have been derived by the experimental upper bounds on the rare CP-violating decays $J/\psi \to K_S^0 K_S^0$ and $\psi(3686) \to K_S^0 K_S^0$, by the BESIII Collaboration \cite{BESIII:2025pnr}:
{\small \begin{align}\label{BESIII}
\mathcal B(J/\psi \to K_S^0 K_S^0) &< 4.7 \times 10^{-9} \, \Rightarrow \, \vert \omega \vert < \Big(4.91 \pm 0.14 \Big) \times 10^{-2}\,, \nonumber \\
\mathcal B(\psi(3686)  \to K_S^0 K_S^0) &< 1.1 \times 10^{-8} \, \Rightarrow \, \vert \omega \vert < \Big(1.44 \pm 0.04 \Big) \times 10^{-2}\,.
\end{align}}
In $B$-systems (Ba-Bar and Belle Collaborations~\cite{BaBar:2014omp}), which are characterised by a negligible difference in width between $B$ and $\overline B$ mesons, one may use the following parametrization for the $\omega$ effect, incorporating also the parameter $\theta$ of conventional CPT Violation ({\it i.e.} with a well-defined CPT generator, {\it e.g.} in Lorentz-Invariance Violation situations) in the initial entangled state $\vert \Psi_0 \rangle$ \cite{Bernabeu:2016kva}:
{\small \begin{align}\label{Bsystemsomega}
\vert \Psi_0 \rangle &\propto \vert B_L \rangle \vert B_H\rangle - \vert B_H \rangle \, \vert B_L\rangle
\nonumber \\
&+ \omega \{ \theta \Big[\vert B_H \rangle \, \vert B_L\rangle + \vert B_L \rangle \, \vert B_H \rangle\Big] + (1 - \theta) \frac{p_L}{p_H} \vert B_H \rangle 
\vert B_H \rangle - (1 + \theta) \frac{p_H}{p_L} \vert B_L\rangle 
\vert B_L \rangle \}\,,
\end{align}}
where $\vert B_{H (L)}\rangle $
are High (Low) mass eigenstates, 
$\widehat H \vert B_{H (L )} \rangle = m_{H (L)} \vert B_{H (L)} \rangle $, 
$\theta = (H_{22} - H_{11})/(m_H - m_L)$, 
with $H_{22}, H_{11}$ the diagonal elements of the Hamiltonian,  
and the parameters $p_{H (L)}$ are defined through:
$\vert B_H \rangle = p_H \vert B_d^0\rangle + q_H \vert \overline{B_d}^0\rangle $, and $ \vert B_L \rangle = p_L \vert B_d^0\rangle - q_L \vert \overline{B_d}^0\rangle$, in the usual notation for B-systems. 
Exploring the equal-sign dilepton asymmetry in such systems (Ba-Bar), one can obtain:\cite{Alvarez:2006ry} 
$ - 0.0084 < \rm Re\omega < 0.0100 $, 
and:\cite{Bernabeu:2016kva}
$\rm Im \theta = \Big(0.99 \pm 1.98\Big)  \times 10^{-2}$, $\rm Im \omega = \pm \Big(6.40 \pm 2.80\Big) \times 10^{-2}$ (the quality of data is not great, hence the above are interpreted only as bounds).

We also remark that the EPR modifications due to the aforementioned spin-system approach may find interesting applications in quantum information and particle cryptography and tomography, topics discussed extensively in this conference (see Yu Shi's and Tao Han's talks). 

\vspace{-0.2cm}\vspace{-0.2cm}

\section{Quantum-entangled gravitons in Rotating Black Holes: $\omega$-effect analogues?}\label{sec:superrad}

\begin{figure}[ht]
    \centering
    \includegraphics[width=0.5\linewidth]{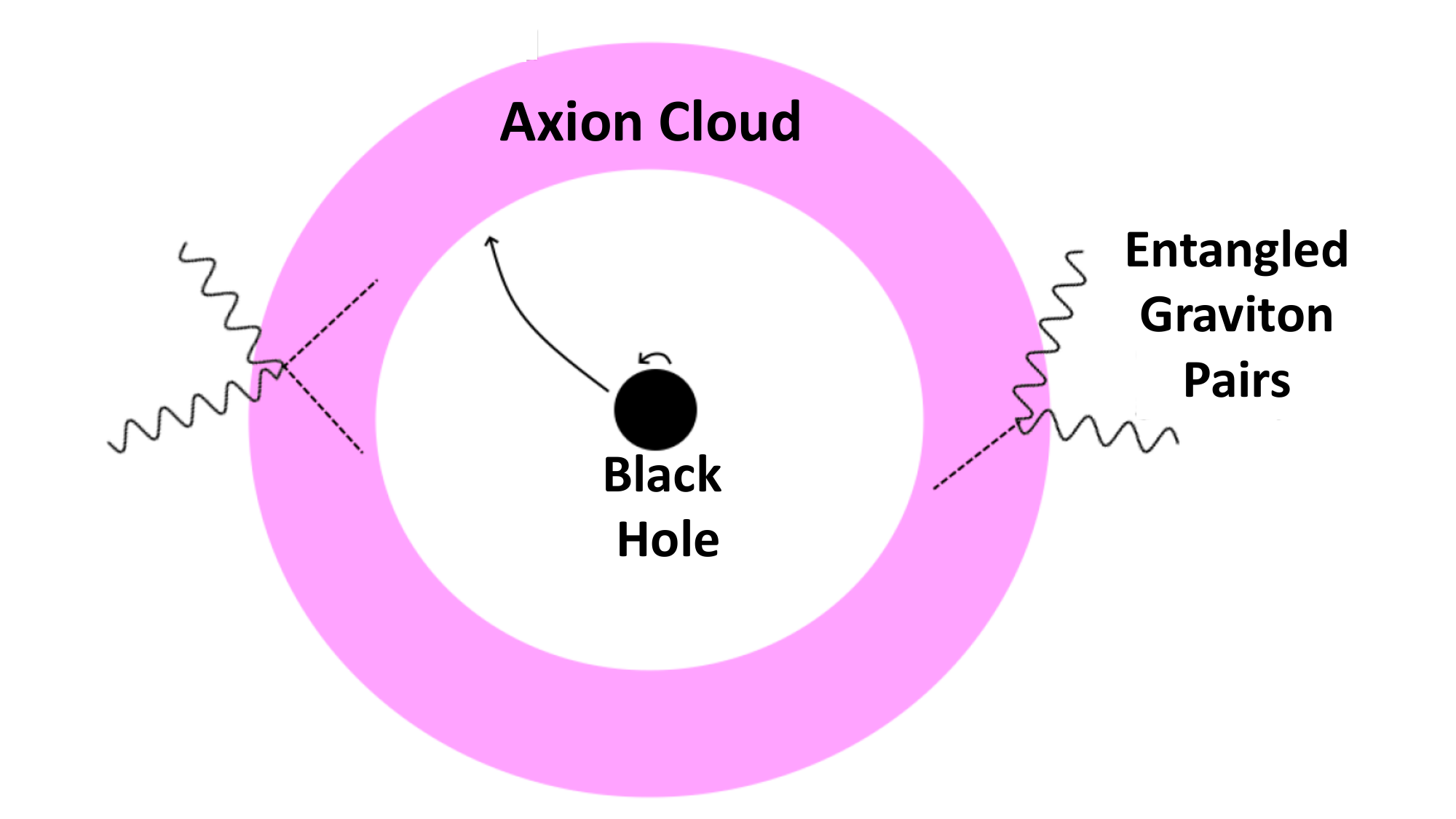}
    \caption{The superradiant axionic cloud (shaded coloured ring) around a rotating black hole (BH) (dark central blob). 
    The processes of the production of squeezed entangled graviton (wavy lines) polarization states from ALPs (dashed lines) are also indicated. Picture adapted from Ref.~\protect\cite{Dorlissuperr}.}
    \label{fig:cloud}
\end{figure}

Rotating (Kerr-type~\cite{Kerr:1963ud}) black holes (BH), of astrophysical interest, 
can support, in the exterior regions of their outer horizons, condensate-like clouds of massive axion-like particles (ALPs) ({\it c.f.} fig.~\ref{fig:cloud}), which can lead to superradiant instabilities,\cite{BritoCardoso} due to the existence of modes with complex frequencies with positive 
imaginary parts. Such instabilities amplify incoming gravitational waves, through the extraction of energy and angular momentum from the BH. In the presence of the massive ALPs, described by a pseudoscalar $b(x)$ field of mass $\mu_b$, in the Kerr geometry, the effective gravitational theory is a Chern-Simons (CS) gravity~\cite{Jackiw:2003pm,Alexander:2009tp}:\footnote{Our conventions and definitions used throughout this work are those of Ref.~\cite{Dorlissuperr}: system of units $\hbar=c=1$, signature of metric $(-, +,+,+ )$, Riemann Curvature tensor:
$R^\lambda_{\,\,\,\,\mu \nu \sigma} = \partial_\nu \, \Gamma^\lambda_{\,\,\mu\sigma} + \Gamma^\rho_{\,\, \mu\sigma} \, \Gamma^\lambda_{\,\, \rho\nu} - (\nu \leftrightarrow \sigma)$, Ricci tensor $R_{\mu\nu} = R^\lambda_{\,\,\,\,\mu \lambda \nu}$, and Ricci scalar $R = R_{\mu\nu}g^{\mu\nu}$. Greek indices are (3+1)-dimensional spacetime ones, taking on values $0, \dots 3.$}
\begin{equation}
	S=\int d^4x \,\sqrt{-g}\,  \left[\frac{R}{2\kappa^2}-\frac{1}{2}(\partial_\mu b)(\partial^\mu b) -\frac{1}{2}\mu^{2}_b \  b^2 - A\, b\,R_{CS} \right] \ ,
\label{eq:Action}
\end{equation}
where  $\kappa=\sqrt{8\pi G}=M_{\rm Pl}^{-1}$ is the inverse of the reduced Planck mass, and $G$ is Newton's gravitational constant. The quantity 
$R_{CS}= \frac{1}{2}R^{\mu}_{\,\,\,\nu\rho\sigma}\widetilde{R}^{\nu\,\,\,\,\rho\sigma}_{\,\,\,\mu}$, with the symbol $\widetilde{(\dots)}$ denoting the dual of the Riemann tensor, defined as
$\widetilde{R}_{\alpha\beta\gamma\delta}=\frac{1}{2}R_{\alpha\beta}^{\,\,\,\,\,\,\,\,\rho\sigma}\varepsilon_{\rho\sigma\gamma\delta}$, where $\varepsilon_{\rho\sigma\gamma\delta}$ is the covariant Levi-Civita symbol, is the gravitational CS anomaly (gCS), which is a total derivative term, having no classical origin and being non zero in Kerr spacetimes, in the presence of ALPs; $A$ is a coupling parameter, of mass dimension $\left[A\right]=-1$. In case the model is embedded in string theory, which is also characterised by a plethora of ALPs~\cite{Svrcek:2006yi}, then~\cite{Duncan:1992vz} 
$A\sim 10^{-2}M_{\rm Pl}/M_s$.

On expanding the effective action \eqref{eq:Action} up to second order in weak tensor perturbations $h_{\mu\nu}$ 
around a fixed spacetime background, which, for the purposes of studying the production of entangled graviton states deep in the axionic cloud \cite{Dorlissuperr}, far away from the exterior BH horizon, can be taken to be the Miknowski spacetime $\eta_{\mu\nu}$: $g_{\mu\nu} = \eta_{\mu\mu} + \kappa \, h_{\mu\nu}$, and quantizing $h_{\mu\nu}$, one can demonstrate the production of non-separable graviton (left(L), right(R)) polarization states, implying entanglement~\cite{Law:2000hyw}. The fusion of two ALPs, which characterises the GR non-anomalous terms in the effective action \eqref{eq:Action} upon the weak-graviton-$h$-expansion, leads to an entangled state of the form \cite{Dorlissuperr}: 
\begin{equation}\label{EPRGR}
    \vert \Psi_{GR}\rangle=\frac{1}{2}\left(\mathcal{G}^{(GR)}_{(R,\vec{k})(L,\vec{k}^\prime)}\vert RL\rangle+\mathcal{G}^{(GR)}_{(L,\vec{k})(R,\vec{k}^\prime)}\vert LR\rangle+\mathcal{G}^{(GR)}_{(L,\vec{k})(L,\vec{k}^\prime)}\vert LL\rangle+\mathcal{G}^{(GR)}_{(R,\vec{k})(R,\vec{k}^\prime)}\vert RR\rangle        \right)\ ,
\end{equation}
with the opposite polarization correlations being enhanced, given that
\begin{equation}\label{EPRGR1}
    \mathcal{G}^{(GR)}_{(R,\vec{k})(L,\vec{k}^\prime)}, \,\, \mathcal{G}^{(GR)}_{(L,\vec{k})(R,\vec{k}^\prime)} \, \gg \, \mathcal{G}^{(GR)}_{(L,\vec{k})(L,\vec{k}^\prime)}, \,\, \mathcal{G}^{(GR)}_{(R,\vec{k})(R,\vec{k}^\prime)}\ .
\end{equation}
The quantities $\mathcal G^{GR}$ are defined in Ref.~\cite{Dorlissuperr}, where it was also shown that: $\mathcal{G}^{(GR)}_{(R,\vec{k})(L,\vec{k}^\prime)} = \mathcal{G}^{(GR)}_{(L,\vec{k})(R,\vec{k}^\prime)} $, which, together with 
\eqref{EPRGR1}, implies the approximate form of the entangled graviton states in the GR framework (in absence of gCS anomalies):
\begin{equation}\label{EPRGRsymm}
    \vert \Psi_{GR}\rangle \simeq \frac{1}{2}\, \mathcal{G}^{(GR)}_{(R,\vec{k})(L,\vec{k}^\prime)} \left(\vert RL\rangle+ \vert LR\rangle \right)\ ,
\end{equation}
that is, the cross-polarization EPR-type~\cite{epr} correlations of the entangled graviton states are approximately symmetric under the interchange of $L \leftrightarrow R$.

The presence of gCS anomalies $R_{CS}$ in \eqref{eq:Action} ``contaminate'' the result \eqref{EPRGRsymm} with antisymmetric entanglement:
\begin{equation}\label{EPRCS}
    \vert \Psi_{CS}\rangle = \frac{1}{2} \, \mathcal{G}^{(CS)}_{(R,\vec{k})(L,\vec{k}^\prime)}\Big( \vert LR\rangle-\vert RL\rangle     \Big)\, ,
\end{equation}
where $\mathcal G^{CS}$ is defined in Ref.~\cite{Dorlissuperr}. This result is reminiscent of, but qualitatively and quantitatively different from, the particle-physics $\omega$ effect ~\cite{Bernabeu:2003ym}, discussed in the previous section, see Eq.~\eqref{omega}. Although there is no ill-defined CPT operator involved in this case, nonetheless the contamination (by the wrong-symmetry entangled-graviton states \eqref{EPRCS} induced by the gCS anomaly) of the GR-induced correlator \eqref{EPRGRsymm} 
is reminiscent of the $\omega$-effect spirit. The purely-quantum-in-origin gCS anomaly violates formally~\cite{Jackiw:2003pm} the conservation of the stress-energy tensor of the axionic matter, due to energy exchange between the ALP field and the gravitational anomaly. 

As argued in \cite{Dorlissuperr}, significant graviton squeezing may be produced for sufficiently long-lived axion clouds from GR contributions in the effective action, while the contribution from gCS anomalies is in general comparatively suppressed. Such effects may be potentially observable in future interferometers~\cite{Arun2022-oe}.

\vspace{-0.2cm}

\section{Conclusions and Outlook}\label{sec:concl}

In the first part of the talk, we reviewed characteristic situations where QG may affect EPR correlations of particles. Specifically, we have discussed potential modifications of EPR correlations in entangled-neutral-meson (in particular Kaon) factories, due to an ill-definition of the CPT-symmetry quantum generator 
in QG models with decoherening environments, consisting of degrees of freedom not accessible to low-energy observers via scattering. This effect is termed ``$\omega$-effect''.

In the second part, we have discussed the emergence of two-mode squeezed graviton states from Axion-like particles in condensate-like clouds of rotating Black Holes. 
Such situations also lead to the appearance of gravitational CS anomalous terms, which, like the $\omega$-effect, affect the pertinent EPR correlations of the quantum-entangled (squeezed) graviton polarization (left, right) states. 

An important future research is to look for multi-graviton entanglement, in analogy with multipartite entanglement in quantum optics, in formalisms where one goes beyond the quadratic-in-weak-graviton approximation in the respective gravitational effective actions. 

\vspace{-0.5cm} 
\section*{Acknowledgements}

NEM wishes to thank the organisers of the WQC Workshop on {\it Quantum Entanglement of High Energy Particles} (Wilczek Quantum Center, USTC Shanghai Institute for Advanced Studies, Shanghai (China), July 19-23 2025) for the invitation to give a plenary talk in this excellent, and thought stimulating, workshop and for their support. He also thanks B.~Hiesmayr for discussions. This work is partially supported by the UK Science and Technology Facilities Council (STFC) under the research grant  ST/X000753/1.

\vspace{-0.4cm}

\bibliographystyle{ws-ijmpa}

\bibliography{QGentangl}

\end{document}